\documentclass[aps,prc,groupedaddress,showpacs,twocolumn,floatfix]{revtex4}

\usepackage{graphicx}
\usepackage[dvips]{color}
\usepackage{colordvi}
\usepackage{latexsym}

\def\gtorder{\mathrel{\raise.3ex\hbox{$>$}\mkern-14mu
 \lower0.6ex\hbox{$\sim$}}}
\def\ltorder{\mathrel{\raise.3ex\hbox{$<$}\mkern-14mu
 \lower0.6ex\hbox{$\sim$}}}

\def\gep{G_{Ep}}
\def\gmp{G_{Mp}}
\def\gen{G_{En}}
\def\gmn{G_{Mn}}
\def\ges{G_{Es}}
\def\gms{G_{Ms}}
\def\gaz{G_A^Z}
\def\apv{A_{PV}}
\def\ans{A_{PV}^{S=0}}
\def\sigr{\sigma_{\mathrm{red}}}

\def\etal{{\it et al.}}
\def\beq{\begin{equation}}
\def\eeq{\end{equation}}

\begin{document}

\title{Precise determination of low-Q nucleon electromagnetic form factors and
their impact on parity-violating $e$--$p$ elastic scattering}

\author{John Arrington}
\affiliation{Argonne National Laboratory, Argonne, IL, USA}

\author{Ingo Sick}
\affiliation{Dept.~f\"{u}r Physik und Astronomie, Universit\"{a}t Basel,
Basel, Switzerland}

\date{\today}

\begin{abstract}

The extraction of the strangeness form factors from parity violating elastic
electron-proton scattering is sensitive to the electromagnetic form factors at
low $Q^2$.  We provide parameterizations for the form factors and
uncertainties, including the effects of two-photon exchange corrections to the
extracted EM form factors.  We study effect of the correlations between
different form factors, in particular as they impact the parity violating
asymmetry and the extraction of the strangeness form factors.  We provide a
prescription to extract the strangeness form factors from the asymmetry that
provides an excellent approximation of the full two-photon correction.  The
corrected form factors are also appropriate as input for other low-$Q$
analyses, although the effects of correlations and two-photon exchange
corrections may be different.

\end{abstract}

\maketitle

\section{Introduction}

The parity-violating (PV) asymmetry in elastic scattering of polarized
electrons from unpolarized protons can be used to extract information on the
strangeness contribution to the proton form factors~\cite{cahn78, mckeown89,
beck89}.  Because the electromagnetic (EM) coupling is proportional to the
quark charge-squared, scattering from the proton is strongly dominated by
interaction with the up quarks. Electron--neutron scattering provides a
different relative weighting of the up and down quark distributions, allowing
one to study the difference between up and down quark contributions to the nucleon
form factor under the assumption that the up-quark distribution in the proton
is identical to the down quark distribution in the neutron, and neglecting
heavier quarks. Because the parity violating cross section comes from
interference between photon and $Z$ exchange, the quark flavors have a
different weighting in the interaction, allowing separation of up, down, and
strange contributions to the form factors by combining proton and neutron
electromagnetic form factors and parity-violating $e$--$p$ scattering. 
However, the small contribution of the strange quarks to the parity-violating
asymmetry requires precise knowledge of the contributions from the up and down
quarks before one is able to achieve sensitivity to the strange
quark contributions.

The parity-violating asymmetry arises due to interference between
photon exchange and $Z$ exchange, and in the Born
approximation is given by~\cite{afanasev05b} 
\beq
\label{eq:born}
\apv^{Born} = - \frac{G_F Q^2}{4 \pi \alpha \sqrt{2}}
\frac{A_E + A_M + A_A}{(\tau \gmp^2 + \varepsilon \gep^2)},
\eeq
where $G_F$ is the Fermi constant, $\alpha$ is the fine structure constant,
and $Q^2$ is the four-momentum transfer squared.  The individual asymmetry
terms can be written in terms of the proton's EM vector form factors, $\gep$
and $\gmp$, and the proton's neutral weak vector and axial form factors,
$\gep^Z$, $\gmp^Z$, and $\gaz$:
\begin{eqnarray*}
A_E = \varepsilon \gep \gep^Z,~~~A_M = \tau \gmp \gmp^Z, \\
A_A = (1 - 4 \sin^2{\theta_W}) \varepsilon^\prime \gmp \gaz,
\end{eqnarray*}
where $\theta_e$ is the electron scattering angle,
$\tau = Q^2/(4 M_p^2)$, $\varepsilon^{-1}=(1+2(1+\tau)\tan^2{\theta_e})$,
$\theta_W$ is the weak mixing angle, and $\varepsilon^\prime =
\sqrt{\tau(1+\tau)(1-\varepsilon^2)}$.

In the standard model and with the assumption of isospin symmetry, the weak
form factors can be expressed in terms of the
proton and neutron EM form factors and the strangeness contribution to the
nucleon EM form factors, $\ges$ and $\gms$, neglecting contributions from
heavier quarks~\cite{beck89}.  Making this substitution, and removing
the common factor $A_0 = -(G_F Q^2)/(4 \pi \alpha \sqrt{2})$, $\apv$ contains
the terms
\begin{eqnarray}
(1 - 4 \sin^2{\theta_W}),~~~~~~~ \label{eq:term1} \\
\frac{-\varepsilon \gep \gen} {\sigr},~~~~~~~ \label{eq:term2} \\
\frac{-\tau \gmp \gmn} {\sigr},~~~~~~~ \label{eq:term3} \\
\frac{-\varepsilon^\prime (1 - 4 \sin^2{\theta_W}) \gmp \gaz}{\sigr}, \label{eq:term4}
\end{eqnarray}
which depend only on quantities that are measured or which can be reliably 
estimated, and one final term,
\beq
\label{eq:strange}
\frac{\varepsilon \gep \ges+ \tau \gmp \gms}{\sigr},
\eeq
which contains the unknown quantities of interest: $\ges$ and $\gms$.  In the
above expressions, we have written the denominator in terms of the $e$--$p$
reduced cross section, $\sigr = \tau \gmp^2 + \varepsilon \gep^2$.

To extract the strangeness-containing term, the best known values and
uncertainties for the other terms are needed. We present an analysis of the
{\em world} $e$--$p$ and $e$--$n$ scattering data to determine the nucleon
form factors, the $e$--$p$ reduced cross section (the denominator of
Eqs.~\ref{eq:term2}-\ref{eq:strange}), and their uncertainties. We also study
the impact of the \textit{correlations} between the different form factors as
well as the effect of two-photon exchange (TPE) corrections on the extraction
of $\ges$ and $\gms$. Electroweak radiative corrections have been
calculated~\cite{musolf94, erler04}, and their uncertainties do not generally
limit the extraction of the strangeness contributions.

The extracted form factors and uncertainties are also appropriate for use in
the analysis of other high precision, low $Q^2$ experiments.  However, the
analysis of correlations and TPE exchange effects presented in this paper is
aimed specifically at parity-violating elastic electron-proton scattering.
Care must be taken in using these fits in analysis of other experiments, as it
is necessary to determine if the analysis requires the Born form factors or
simply needs the form factors as a parameterization of the elastic cross
section.  When the cross section is required, using the Born form factors
requires making an explicit correction for TPE effects~\cite{arrington04a}. 
For other cases, such as the extraction of the axial form factor from neutrino
scattering~\cite{budd03}, determining corrections to hyperfine splitting in
hydrogen~\cite{brodsky04b}, or determining the Bethe-Heitler term in the
analysis of DVCS measurements, one needs to carefully consider whether TPE
corrections are needed and to which degree they are different from the ones
needed for the unpolarized cross section.

\section{Analysis of low-$Q$ data}

\subsection{Proton form factors}\label{sec:proton}

A fit to the world e-p cross section data at very low momentum transfer
has been described in~\cite{sick03}.  This fit uses a Continued Fraction (CF)
expansion,
\beq
 G_{CF}(Q) = \frac{1}{\displaystyle 1
                    + \frac{b_1 Q^2 \hfill }{\displaystyle 1
		       + \frac{b_2 Q^2 \hfill }{\displaystyle 1
		          + \cdots}}},
\label{eq:cf}
\eeq
of $\gep$ and $\gmp$ most suitable for the lower momentum transfers, and
extends up to $Q=\sqrt{Q^2} \approx 0.8$~GeV/c.  Note that these fits should
only be used in the quoted range of $Q$ values.  The analysis includes the
effects of Coulomb distortion which, contrary to common belief, are {\em not}
negligible~\cite{arrington04c}. The effect of two-photon exchange {\em beyond}
Coulomb distortion, which includes only the exchange of an additional soft
photon, has also been studied~\cite{blunden05a,borisyuk06}. In our main
analysis, we will correct the proton cross sections for Coulomb distortion,
though we also provide parameterizations using the full calculation for TPE
effects and discuss the impact of TPE corrections on the neutron form factors.

Here, we extend this fit to higher momentum transfers, up to $Q = 1.2$~GeV/c,
such as to sufficiently bracket the kinematics covered by the different PV
experiments.  The approach taken is identical to the one employed
in~\cite{sick03}:  We start from the {\em world} cross sections for $e$--$p$
scattering~\cite{bumiller61, janssens66, borkowski74, borkowski75, simon80,
simon81, albrecht66, bartel66, frerejacque66, albrecht67, bartel67, bartel73,
ganichot72, kirk73, murphy74, berger71, bartel70}, apply the Coulomb
corrections according to~\cite{sick98,arrington04c} and fit the cross sections
with CF parameterizations of both $\gep$ and $\gmp$. The
longitudinal/transverse (L/T)-separation is done implicitly by the fit.

This approach allows one to keep track of the random and systematic
uncertainties of the data and propagate them to the final quantities of
interest. From the resulting CF parameters and the error matrix of the fit one
can calculate the values of $\gep$ and $\gmp$  together with their random
error for any desired value of $Q$. To obtain the systematic error of the
proton form factors, each data set is changed by the quoted systematic
uncertainty, the world set is refit, and the change in $\gep$ and $\gmp$
calculated. These changes are added up quadratically for all data sets,
yielding the systematic uncertainty on $\gep$ and $\gmp$. This is usually the
dominant error. The total error is obtained by adding quadratically the random
and the systematic errors, a procedure that should be applicable given the
large number of data sets used.

\begin{table}
\caption{Fit parameters for the low-Q form factors, valid up to $Q=1$~GeV/c,
using the CF parametrization of Eq.~\ref{eq:cf} (with $Q^2$ in (GeV/c)$^2$) for
$\gep$, $\gmp/\mu_p$, $\gmn/\mu_n$, and the parameterization of
Eq.~\ref{eq:gen} for $\gen$.  The proton data are corrected for Coulomb
distortion.
\label{tab:fit}}
\begin{ruledtabular}
\begin{tabular}{l|ccccc}
	& $b_1$ & $b_2$	& $b_3$      & $b_4$	& $b_5$	\\
\hline
$\gep$		& 3.440 & --0.178 & --1.212 &  1.176 & --0.284 \\
$\gmp / \mu_p$	& 3.173 & --0.314 & --1.165 &  5.619 & --1.087 \\
$\gen$		& 0.977 & --20.82 &   22.02 &    -   &    -    \\
$\gmn / \mu_n$	& 3.297 & --0.258 &   0.001 &    -   &    -    \\
\end{tabular}
\end{ruledtabular}
\end{table}

\begin{figure}[htb]
\includegraphics[scale=0.49,clip]{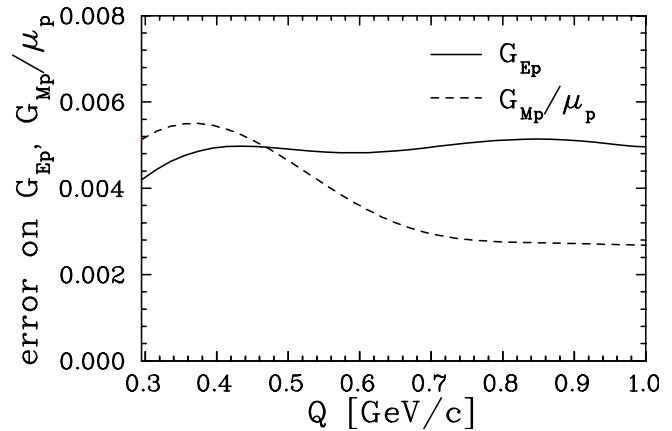}
\caption{Uncertainties in the fits for $\gep$ (solid) and $\gmp/\mu_p$
(dashed). The uncertainty is the random and systematic uncertainties, combined
in quadrature.
\label{fig:proton}}
\end{figure}

The fit to the proton cross sections, after correcting for Coulomb distortion,
yields the coefficients given in Table~\ref{tab:fit}. The fits are valid for
$Q$ from 0.3 to 1.0 GeV/c.  The uncertainty for $\gep$ ($\gmp / \mu_p$), is
given by the solid (dashed) line in Fig.~\ref{fig:proton}.
At low $Q$, the error bar on
$\gmp$ is larger than the one on $\gep$ as the data base is less complete,
although in the low-$Q$ region there are two data sets with measurements at
180$^\circ$~\cite{frerejacque66, ganichot72}.

\subsection{Neutron form factors}\label{sec:neutron}

The most precise data for the neutron magnetic form factor $\gmn$ come from
measurements of the ratio of $^2$H($e,e'n$) to $^2$H($e,e'p$)~\cite{kubon02,
anklin98}.  The value of $\gmn$ is extracted from the neutron cross section,
which is determined from the combination of the neutron to proton ratio in
deuterium and the (free) proton elastic cross section.  We also include
measurements from the asymmetry on polarized $^3$He~\cite{anderson06}, which
are of somewhat lower precision, and data points for $Q < 1.3$~GeV/c from
Ref.~\cite{rinat04}.  The high $Q$ points have larger uncertainties, and
are outside the range of validity of the fit, but are included to avoid
``extreme'' behavior for $Q < 1$~GeV/c.  We fit these data to a 3rd order
CF expansion, and the parameters are shown in Tab.~\ref{tab:fit}.  The random
and systematic uncertainties of $\gmn$ have been estimated in
Ref.~\cite{kubon02}.  In the range of momentum transfer of interest here, they
are approximately 1.5\%, roughly independent of $Q$.

The value of the neutron charge form factor, $\gen$, is obtained by fitting all
data presently available from polarization-transfer
experiments~\cite{passchier99, herberg99, ostrick99, becker99, zhu01,
bermuth03, madey03, warren04, glazier05}.  Care has been taken to employ the
most recent values, as some of the experimental $\gen$'s published early on
did not contain the best corrections for FSI and MEC (or no corrections at
all).  To study the uncertainty due to FSI and MEC corrections, an additional
uncertainty equal to 30\% of the calculated correction was added in quadrature
with the experimental uncertainties. Including this additional uncertainty in
the extraction of $\gen$ has little effect on the fit or the uncertainties.
Also included in the fit are the $\gen$-values determined from the deuteron C2
form factor~\cite{schiavilla01} and the slope of $\gen$ at $Q=0$, known from
$n$--$e$ scattering. For $\gen$ the error bars of the published data contain a
mix of random and systematic uncertainties. This mix is difficult to take
apart, and therefore no distinction between random and systematic errors is
made here.

The fits of $\gen$ are done using a modified 3-parameter CF expansion:
\beq
  \gen(Q) = 0.484 \cdot Q^2 \cdot G_{CF},
  \label{eq:gen}
\eeq
with $Q^2$ in (GeV/c)$^2$.  The constant value in front fixes the slope at
$Q^2=0$ to match the measured $rms$-radius squared value of
--0.113~fm$^2$~\cite{koester95}.  The fit parameters are given in
Tab.~\ref{tab:fit}, and the fit is valid up to $Q=1$~GeV/c.

\begin{figure}[htb]
\includegraphics[scale=0.49,clip]{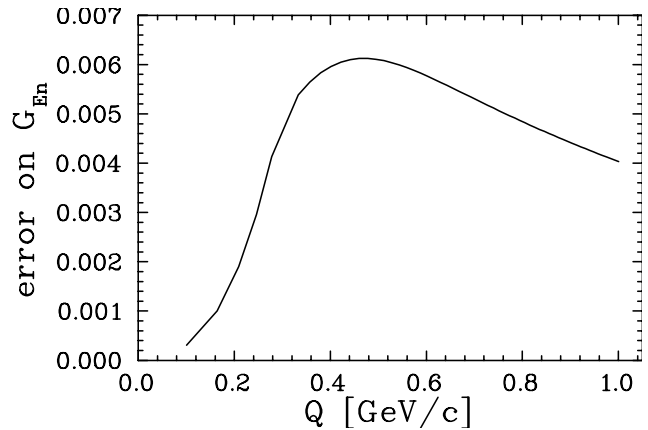}
\caption{Total uncertainty for the fit to $\gen$.
\label{fig:gen}}
\end{figure}

The error matrix is used to compute the error of $\gen$ at any desired value
of $Q$.  Fits using a functional form similar to the Galster
fit~\cite{galster71} are systematically below our fit, while fits attempting
to include an explicit pion cloud contribution~\cite{friedrich03} lie above
our fit.  We performed fits using different fit functions and took the fit
dependence as an additional contribution to the uncertainty in $\gen$.  This
is the dominant source of the uncertainty below $Q \approx 0.3$~GeV/c, where no
direct measurements exist.  The final estimated uncertainty on
$\gen$ is shown in Fig.~\ref{fig:gen}.  The figure shows the absolute
uncertainty on $\gen$; the relative uncertainty is well parameterized by
taking the minimum of ($5.2 + 12.6 \cdot Q^2$)\%, which fits the curve below
$Q=0.3$~GeV/c, and ($9.2 + 11 \cdot \exp{(-Q^2/0.19)})$\%, which fits above
$Q=0.3$~GeV/c.

\subsection{Cross Section}\label{sec:sigr}

In the Born approximation, $\apv$ is given by Eq.~\ref{eq:born}.  The
inclusion of two-photon exchange terms leads to the replacement of the Born
form factors $\gep(Q^2)$ and $\gmp(Q^2)$ with generalized form factors that
depend on both $\varepsilon$ and $Q^2$, as well as introducing two new terms,
$A_M'$ and $A_A'$~\cite{afanasev05b}.  Given a complete calculation of the
two-photon exchange correction, one can extract the Born form factors by
correcting the Rosenbluth and polarization extractions for TPE effects, and
then applying the TPE corrections to $\apv$. However, the TPE corrections to
the denominator of Eq.~\ref{eq:born} are identical to the corrections to the
$e$--$p$ unpolarized cross section. So rather than correcting the unpolarized
cross section measurements for TPE and then re-applying TPE correction to
evaluate $\sigr$, one can make a model-independent evaluation of $\sigr$ by
taking a fit to the TPE-uncorrected $e$--$p$ cross section.  Thus, we also
provide a fit to the measured $e$--$p$ cross section, without applying any
kind of TPE correction.

The procedure is identical to the extraction of the proton form factors,
except that a global fit is performed to the {\em uncorrected} cross
sections.  The reduced cross section is fit to the the form $\sigr = \tau
F_m(Q^2) + \varepsilon F_e(Q^2)$, such that in the Born approximation, $F_m =
\gmp$ and $F_e = \gep$.  While the fit to the TPE-uncorrected data can also
have an $\varepsilon$ dependence, a global analysis of the $\varepsilon$
dependence of $\sigma_{ep}$ indicates that that deviations from linearity are
extremely small~\cite{tvaskis06}.  Table~\ref{tab:sigrfit} gives the parameters
for the fit to the uncorrected cross sections.  This fit is appropriate both
for the reduced cross section term in the evaluation of $\apv$, but also as
a parameterization of the elastic cross section with the TPE corrections
absorbed into the fit function.  It is therefore useful as a low $Q$ model
of the elastic cross section if one does not wish to explicitly treat the TPE
corrections for unpolarized $e$--$p$ scattering.

\begin{table}
\caption{Fit parameters for the low-Q $e$--$p$ cross section, neglecting TPE
corrections.
\label{tab:sigrfit}}
\begin{ruledtabular}
\begin{tabular}{c|ccccc}
	& $b_1$ & $b_2$	& $b_3$      & $b_4$	& $b_5$	\\
\hline
$F_e$		& 3.366 & --0.189 & --1.263 &  1.351 & --0.301 \\
$F_m / \mu_p$	& 3.205 & --0.318 & --1.228 &  5.619 & --1.116 \\
\end{tabular}
\end{ruledtabular}
\end{table}

\begin{figure}[htb]
\includegraphics[scale=0.49,angle=0,clip]{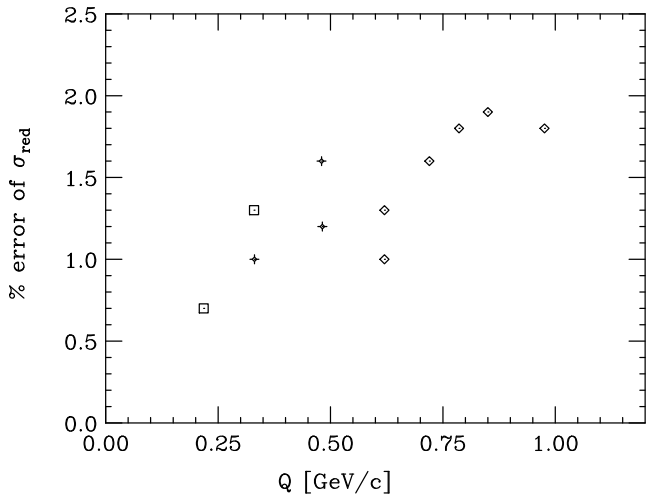}
\caption{Total uncertainty for the fit to the TPE-uncorrected value of $\sigr$
at the kinematics of several past and planned measurements.
The kinematics of forward angle are shown for the JLab ($\Diamond$), while
backward angle kinematics are shown for Bates ($\Box$) measurements.  The
Mainz ($+$) points are for forward angle measurements at low $Q$, and backward
angle measurements at higher $Q$.
\label{fig:sigr}}
\end{figure}

The uncertainties in $F_e$ and $F_m$ are essentially identical to those
of $\gep$ and $\gmp$.  However, the calculation of the uncertainty in $\sigr$
requires special care.  The values of $F_e$ and $F_m$ are strongly correlated
as they result from the (implicitly made) L/T-separation of the cross
sections.  The uncertainty on the cross section thus is smaller than the one
one would obtain by combining the errors in $F_e$ and $F_m$.  The
error matrix is used to evaluate the random uncertainties, while the
systematic uncertainty is taken as the combined effect of individually varying
the normalization of each data set.  It is not possible to provide a simple
parameterization for the cross section uncertainty at all $\varepsilon$ and
$Q$ values.  The cross section uncertainties at kinematics corresponding to a
variety of PV experiments are given in Figure~\ref{fig:sigr}.  For the forward
angle measurements, the cross section uncertainty is a relatively simple
function of $Q$, so the uncertainty for other large-$\varepsilon$ can be
estimated from the small angle data in Fig.~\ref{fig:sigr}.  The $Q$ dependence
is more complicated for large angle, and we have tried to include the complete
set of planned measurements.

\subsection{Two-photon exchange beyond Coulomb distortion}\label{sec:tpevcoul}

In the extractions of the proton form factors described above, we have applied
Coulomb distortion corrections~\cite{sick98,arrington04c} to the $e$--$p$
scattering data, and no correction to the $e$--$n$ data. Coulomb distortion
takes into account the effect of a second soft photon, but does not include the
contribution from a second hard photon.  At low $Q$, the difference between
Coulomb distortion and full TPE is small~\cite{blunden05b}.

To gauge the effect of the full two-photon corrections, we perform another
extraction of the proton form factors, after correcting $\sigma_{ep}$ using
the TPE calculation of Blunden \etal~\cite{blunden05a}, rather than the
Coulomb distortion correction. These authors calculate the contribution of
the exchange of a second photon, soft or hard, restricted to the case for the
intermediary state being a proton in its ground state.  This calculation 
includes only the unexcited proton in the intermediate state; the contribution
from excited intermediate states is neglected.  This calculation explains most
but not all of the discrepancy between Rosenbluth and Polarization
measurements above $Q^2$=2~(GeV/c)$^2$, but appears to be sufficient at lower
$Q^2$ values. A calculation including an intermediate
$\Delta$~\cite{kondratyuk05} also indicates that this contribution is
important at $Q^2$=2--3~(GeV/c)$^2$, but provides only a small modification
below 1~(GeV/c)$^2$.

\begin{table}
\caption{Fit parameters for the low-Q proton form factors, using the two-photon
exchange correction from Ref.~\cite{blunden05a}.
\label{tab:fit2}}
\begin{ruledtabular}
\begin{tabular}{c|ccccc}
	& $b_1$ & $b_2$	& $b_3$      & $b_4$	& $b_5$	\\
\hline
$\gep$		& 3.478 & --0.140 & --1.311 &  1.128 & --0.233 \\
$\gmp / \mu_p$	& 3.224 & --0.313 & --0.868 &  4.278 & --1.102 \\
\end{tabular}
\end{ruledtabular}
\end{table}

Table~\ref{tab:fit2} shows the results of the fit to the proton cross sections
corrected for TPE, as opposed to the Coulomb distortion corrections used for
the fits in Table~\ref{tab:fit}. For this fit, the explicit Coulomb distortion
corrections done above were omitted, as the TPE corrections already contain
the contribution from Coulomb distortion. The fit is valid for $Q$ from 0.3 to
1.0 GeV/c. The uncertainties of $\gep$ and $\gmp$ (random plus systematic added
quadratically) are essentially identical to the fit with Coulomb distortion
(Fig.~\ref{fig:proton}).
While some of the parameters in the fits are noticeably different, the
difference between correcting for Coulomb distortion and TPE on the extracted
form factors is small.  While the size of the Coulomb distortion corrections
can be up to 3\% for $\gep$ and 1\% for $\gmp$, the difference between Coulomb
and full TPE corrections is typically 0.3--0.4\%, and never more than 0.5\% for
$\gep$ and 0.7\% for $\gmp$. This difference is always less then the
uncertainties shown in Fig.~\ref{fig:proton}, and more importantly, is well
within the radiative correction uncertainties assumed in the initial
measurements, which are dominated by the uncertainty in TPE contributions.
While the estimates of the uncertainties in the radiative correction procedure
were clearly underestimates when neglecting TPE corrections, they provide a
reasonable estimate of the uncertainty in the TPE calculation, especially for
these small $Q$ values.

For the neutron, the TPE correction to the cross section as calculated by
Blunden \etal~\cite{blunden05a} is well parameterized at low $Q$ as $\Delta
\sigma / \sigma =  0.8\% \cdot Q^2 \cdot (1-\varepsilon)$, with $Q^2$ in
(GeV/c)$^2$.
For $Q < 1$~GeV/c,this yields a maximum correction go $\gmn$ of 0.4\% at
$Q=1$~GeV/c and large scattering angle.  Since most measurements are at
relatively forward angle, the typical correction is $\ltorder$0.1\%.

The calculated two-photon corrections to the polarization measurements of the
neutron electric form factor are extremely small. It must be noted, however,
that most modern experiments determining $\gen$ measure an asymmetry depending
on the ratio $\gen/\gmn$, so a TPE correction to $\gmn$ propagates into the
extracted value of $\gen$.  As this correction is well below 1\%, the effect
is negligible compared to the experimental uncertainties of these measurements.

Thus, the overall difference between the full TPE correction and the Coulomb
distortion is quite small.  For protons, this different can amount to about
half of the final uncertainty coming from the input form factors, but is
within the radiative correction uncertainty applied in the individual
measurements, and thus is properly accounted for in the final uncertainties.
However, one also has to consider correlations between the uncertainties in
different form factors, which can enhance the effect on the parity violating
asymmetry. This is discussed in Sec.~\ref{sec:correlations}.

\section{Determination of $\apv$}

Given the nucleon electromagnetic form factors, as well as $\theta_W$
and $\gaz$, one can calculate $\ans$, the PV asymmetry for $\ges = \gms =
0$.  One then takes the difference between the measured asymmetry and $\ans$
as the contribution from the unknown term (Eq.~\ref{eq:strange}).  To obtain
a reliable value for the strangeness contribution, one must include radiative
corrections in evaluating $\ans$, and determine the uncertainty in $\ans$
due to uncertainty in the form factor and other terms.  The uncertainty in
$\ans$ is usually determined by varying the individual form factors that
contribute to the asymmetry by their assumed uncertainties.  This approach
ignores two effects which could be significant in these measurements.  First,
it neglects the correlation between the extracted values of the
electromagnetic form factors, which can impact the total uncertainty on
$\ans$. Second, it neglects the impact of two-photon exchange corrections on
$\ans$, as well as their effect on the extracted values of the form factors.
In the following sections, we will study the effect of the correlated
uncertainties between the extractions of the different form factors, and
estimate the size of TPE corrections.  We will present a procedure for
determining the size and uncertainty of the parity-violating asymmetry that
does not require an explicit calculation of TPE, but which minimizes the
uncertainty in extracting the strangeness contributions.  We will compare
this to the result obtained if one ignores both TPE corrections to $\apv$
and the TPE corrections in the extraction of the electromagnetic form factors,
as has been done in all previous extractions of the strangeness contributions.

\subsection{Impact of correlations}\label{sec:correlations}

The effects of these correlations need to be evaluated to obtain an accurate
measure of the uncertainty in $\ans$.  Taking the correlations into account
can noticeably increase or decrease the contribution of the form factors to the
total uncertainty. We examine here the impact of these correlations on the
evaluation of $\ans$, in the region of $0.3 < Q < 1.0$~GeV/c, where such
measurements have been carried out or proposed.

At the $Q$ values of interest here the main issues of concern are the
anti-correlation between $\gep$ and $\gmp$ extracted from Rosenbluth
separation, the correlation between $\gmn$ and $\sigma_{ep}$ ($\sigr$) for data
extracted in measurements of the proton/neutron ratio or the $^3$He
quasielastic asymmetry, and the correlation between $\gen$ and $\gmn$ in
polarization measurements that extract $\gen/\gmn$.  In some cases, it is
difficult to precisely quantify the level of correlation and difficult to
propagate to the value of $\ans$.  The aim of this analysis is to determine
where these correlations can be neglected or treated in some approximate
fashion, and to determine the uncertainty related to these approximations.

\subsubsection{Correlation between $\gep$ and $\gmp$} 

At the low values of $Q$ of interest here, the proton form factors are determined by 
L-T separations, which yield a significant anti-correlation between the
extracted values of $\gep$ and $\gmp$.  This has an impact in determining the
uncertainty in $\sigr$, which appears in the denominator of most of the terms,
as well as introducing a correlation between the errors in the term involving
$\gep$ and the term involving $\gmp$.

The largest effect results from the fact that the uncertainty on $\sigr$ is much
smaller than one obtains by varying $\gep$ and $\gmp$ individually.
By treating $\sigr$ and $\delta \sigr$ as being independent quantities from 
$\gep$ and $\gmp$, we eliminate the overestimate of the uncertainty,
and we better account for TPE effects as well (see Sec.~\ref{sec:sigr}).
When $\sigr$ is extracted directly from the cross sections, the remaining
correlation between $\sigr$ and the individual form factors is very small.

The remaining effect is the correlation between the terms in $\apv$ involving
$\gep$ and $\gmp$ (Eqs.~\ref{eq:term2} and~\ref{eq:term3}).  These terms have
the opposite sign in the final asymmetry, so the anti-correlation between the
values of $\gep$ and $\gmp$ will tend to increase the total uncertainty. For
small $\varepsilon$ values, the term involving $\gep$ is only a few percent of
the total asymmetry, and so its uncertainty has a negligible small effect
($<0.1\%$ of $\ans$ for $\varepsilon<0.05$).  At large $\varepsilon$ values,
this term contributes roughly 20\% the of $\ans$, and so has a greater impact.
 If taken to be 100\% anti-correlated with $\gmp$, the difference at large
$\varepsilon$ grows from 0.1\% of $\ans$ at $Q^2=0.1$~(GeV/c)$^2$ to 0.4\% at
1~(GeV/c)$^2$.  In reality, the effect will be smaller, as the correlation is
not 100\%, especially in the region where $\gmp$ is mainly given by the data
taken at 180$^\circ$. The completed and proposed measurements of $\ans$
typically have $\sim$10\% precision on $\ans$, and never better than 4\%, and
so neglecting this correlation will again have a very small effect on the
final uncertainty. Thus, it is a good approximation to treat $\gep$ and $\gmp$
as uncorrelated, as long as one takes the uncertainty $\delta \sigr$ directly,
rather than calculating $\delta \sigr$ from $\delta \gep$ and $\delta \gmp$.

\subsubsection{Correlation between $\gmn$ and $\sigr$} 

The most precise values of $\gmn$ come from measurements of the ratio
of d(e,e'n) to d(e,e'p).  By normalizing to $\sigma_{ep}$, measured on the
proton, one can extract $\sigma_{en}$ and thus $\gmn$, since the contribution
from $\gen$ is almost negligible.  Measurements utilizing quasielastic
scattering of polarized electrons from polarized $^3$He are essentially
measuring the same quantity~\cite{anderson06}.  The transverse asymmetry for
scattering from the polarized neutron is nearly independent of the neutron
form factors, and so the $^3$He asymmetry is mainly sensitive to the dilution
due to the two (nearly) unpolarized protons, and thus is sensitive
to $\sigma_{ep}/\sigma_{en}$. Therefore, both experiments yield a direct
correlation between the extracted value of $\gmn$ and the value of
$\sigma_{ep}$ used in the analysis.

Because of this correlation, it is important that TPE are treated in a
consistent fashion.  If the proton form factors are corrected for TPE, then
the TPE contributions must be included in calculating $\sigma_{ep}$ as observed
in the $\sigma_{en}/\sigma_{ep}$ measurements.  Because TPE corrections were
neglected in both the extraction of the proton form factors and the calculation
of $\sigma_{ep}$ as used in the $\gmn$ extractions, one obtains a correct
parameterization of the unpolarized $e$--$p$ cross section, as in
Sec.~\ref{sec:sigr}.

The typical uncertainties in $\sigma_{ep}$ at kinematics where $\gmn$ has been
extracted are $\approx$1.4\%, yielding a contribution to the uncertainty in
$\gmn$ of $\approx$0.7\%.  For $Q < 1$~GeV/c, the uncertainty in $\ans$ due to
$\gmn$ is close to half the size of the uncertainty coming from $\sigr$, so a
perfect correlation would have the effect of reducing this contribution to the
uncertainty by roughly a factor of two.  However, the uncertainty on $\sigr$
is usually not the dominant contribution, and so the effect of reducing this
contribution is never more than 0.4\% of $\ans$.  In fact, the effect is even
smaller since the uncertainties are not 100\% correlated between different
extractions of $\gmn$.

\subsubsection{Correlation between $\gen$ and $\gmn$} 

The polarization measurements are sensitive only to the ratio $\gen/\gmn$, and
thus the error in $\gmn$ used in the analysis yields an identical shift
in $\gen$.  However, the contribution to $\delta \gen$ from $\delta \gmn$ is a
very small part (typically 10\%) of the total uncertainty in $\gen$.

\subsection{Two-photon exchange corrections to $\apv$}\label{sec:tpe}

A full calculation of the TPE corrections to $\apv$ requires starting
with the TPE-corrected (i.e. Born) form factors, and then applying the full
TPE corrections for parity-violating scattering to Eq.~\ref{eq:born}.  We have
made fits to the proton and neutron form factors, both with partial TPE
corrections, neglecting the effect of a second hard photon, and with full
TPE corrections.  The full corrections are more model dependent, but are
close to the corrections applied, as discussed in Sec.~\ref{sec:tpevcoul}. The
uncertainties assumed for the radiative corrections are now consistent with
the corrections applied, even for the partial correction.

We apply TPE corrections based on the formalism by Afanasev and
Carlson~\cite{afanasev05b}.  This includes the effect of the two photon
box (and crossed-box) diagrams, but not the effect of the $\gamma$-Z box
diagram, which has been examined (for $Q^2=0$) in Ref.~\cite{marciano84}. It
should be noted that for these corrections, it is \textit{not} sufficient to
apply the correction for only the second soft photon; one must go beyond 
Coulomb distortion.  This is because the Coulomb distortion is a long range
contribution that, to first order, yields a helicity-independent rescaling of
the cross sections, and thus cancels in the evaluation of $\apv$.  So for
$\apv$, one must use the full calculation, including the exchange of a
hard photon.

For convenience, we separate the TPE effects into three categories.  First,
there are two new terms, $A_M'$ and $A_A'$, that appear in the expression for
$\apv$.  Second, the Born form factors that go into the terms $A_E$, $A_M$,
and $A_A$ in Eq.~\ref{eq:born} are replaced with generalized form factors that
depend on both $\varepsilon$ and $Q^2$.  Finally, the TPE correction changes
the unpolarized $e$--$p$ reduced cross section, $\sigr$, which appears as part
of the denominator.  This involves both replacing the Born form factors with
the generalized form factors, and introducing new terms related the new
amplitude that appears when including TPE.

We evaluate the corrections to $\apv$ using the TPE calculations of
Refs.~\cite{borisyuk06, blunden05a}, which give identical results for
the $Q$ range of interest.  The new terms, $A_M'$ and $A_A'$, have 
small contributions, below 0.1\% to $\apv$ for $Q < 1$~GeV/c.  While these
terms become comparable in size to the corrections to $A_A, A_E,$ and $A_M$
at higher $Q$ values, they are negligible at low $Q$.  The effect of the
generalized form factors on $A_E$, $A_M$, and $A_A$ is larger, but still
relatively small; on the order of 1\% for $Q<1$~GeV/c. The final TPE
contribution, the correction to $\sigr$ (the denominator of Eq.~\ref{eq:born})
can be 2--3\% at small $\varepsilon$ values.  Due to cancellation between the
different terms, the combined effect of TPE on $\apv$ is $\ltorder$1\% for $Q
< 1$~GeV/c.

However, the TPE correction to the $\apv^{Born}$ is not the complete story;
one must also take into account the \textit{indirect} impact of TPE
corrections on the extraction of $\ans$.  In the past, TPE contributions were
neglected in extracting the nucleon electromagnetic form factors, and so
the calculated value of $\apv$ is \textit{not} the correct value for
$\apv^{Born}$.  The TPE corrections to $\gep$ and $\gmp$ change $\ans$ by
1--2\% for $Q$ below 1~GeV/c, and as much as 5\% for $Q=2$~GeV/c. The largest
corrections coming for small scattering angles, where the precision of the
completed and planned measurements is the highest, and so one must apply TPE
correction to the extraction of the nucleon electromagnetic form factors, as
done in Section~\ref{sec:proton}.

In summary, the largest effects  are the corrections to the Rosenbluth
extracted values of $\gep$ and $\gmp$, and the application of the TPE effects
to the denominator of Eq.~\ref{eq:born}.  The TPE effect on the denominator
are identical to those in the unpolarized cross section measurements, and so a
model-independent extraction of the denominator can be achieved by using
\textit{uncorrected} $e$--$p$ cross section. We have provided TPE-corrected
fits to the form factors, as well as uncorrected fits to the unpolarized cross
sections, which allow these corrections to be applied without requiring an
explicit calculation of the TPE corrections to $\ans$.  If one neglects the
remaining corrections to the numerator of Eq.~\ref{eq:born}, the result is
within 1\% of the full calculation.  However, a simple linear parameterization
of these remaining terms provides a calculation of $\ans$ that is within 0.2\%
of the full calculation:
\begin{equation}
\apv \rightarrow \apv \cdot [1+(C_0+\varepsilon C_1)]
\label{eq:correction}
\end{equation}
where $C_0 = .013-.022Q$, $C_1 = -.010+.018Q$, with $Q$ in GeV/c.

\begin{figure}[htb]
\includegraphics[scale=0.47,angle=0,clip]{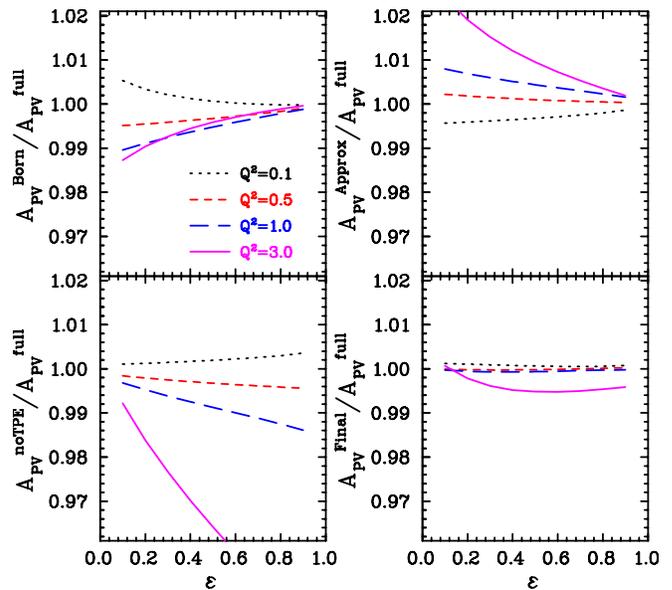}
\caption{Comparisons of different calculations of $\apv$ to the calculation
including the full TPE effects.  Top left plot is $\apv^{Born}$, bottom left
is neglecting TPE in both the extraction of the EM form factors and in
calculating $\apv$, i.e. the procedure used in analyzing previous experimental
results.  The top right is the approximation presented here,
\textit{neglecting} the additional parameterization of the TPE effect on the
numerator of Eq.~\ref{eq:born}, and the bottom right is the final
prescription, including this correction (Eq.~\ref{eq:correction})
\label{fig:pv_tpe}}
\end{figure}

Figure~\ref{fig:pv_tpe} compares various approximations for $\apv$ to the full
calculation explicitly including TPE corrections. The top left panel shows
that the correction to the Born value is small, due to the relatively small
direct TPE contributions, and the cancellation between TPE contributions to
different terms.  The bottom left plot shows the error made when neglecting
TPE corrections in both the calculation of $\apv$ \textit{and} the extraction
of the EM form factors. The right hand plots show the approximation discussed
in this paper, neglecting the additional correction due to the effect on the
numerator in Eq.~\ref{eq:born} (top figure), and including the parameterization
of this correction from Eq.~\ref{eq:correction} (bottom figure).

\subsection{Extension to larger $Q$}

While corrections to individual terms in Eq.~\ref{eq:born} can be at the
1--2\% level, and additional corrections due to TPE effects in the extraction
of the Born EM form factors can be even larger, significant cancellation
between different terms yields a total correction that is typically below 1\%
for $Q < 1$~GeV/c.  After applying the TPE corrections as discussed above, the
uncertainties in the TPE corrections for $Q<1$~GeV/c are dominated by the
uncertainty in extracting the TPE-corrected form factors.  This uncertainty is
taken into account in the typical 1.5\% uncertainty assumed for radiative
corrections, and thus no additional uncertainty need be applied.

At larger $Q$, these corrections grow significantly, as shown in
Fig.~\ref{fig:pv_tpe}, and the total error made in neglecting TPE corrections
can reach 10\% by $Q=2$~GeV/c.  The procedure described here is provides a
correction good to 0.2\% up to $Q=1$~GeV/c, and 1\% up to 2~GeV/c. At higher $Q$,
the corrections become even larger, and the calculation of TPE corrections
becomes less reliable.  An estimate of the contributions from an intermediate
$\Delta$ in the box diagram~\cite{kondratyuk05} indicates that this
contribution is less than 0.3\% for $Q^2<1$~(GeV/c)$^2$, while at
$Q^2=3$~(GeV/c)$^2$, the contribution is as large as 2\%, and is significantly
more model dependent.

\section{Summary of the procedure}\label{sec:uncertainties}

The final prescription involves evaluating the terms in Eq.~\ref{eq:born},
using TPE-corrected fits for the nucleon form factors in the terms $A_E$,
$A_M$, and $A_A$, the \textit{TPE-uncorrected} fits to $\sigr$ for the
denominator, and applying the correction from Eq.~\ref{eq:correction} to
account for the TPE corrections for the terms in the numerator of
Eq.~\ref{eq:born} and the additional terms $A_M'$ and
$A_A'$~\cite{afanasev05b}.  Without this final correction, the approximation
is valid to better than 1\% for $Q$ values from 0.3-1.0~GeV/c, and better than
0.5\% except for $Q \approx 1$~GeV/c and $\varepsilon < 0.5$.  With this correction, 
the approximation is good to 0.2\%.

To get the overall error of the term  Eq.~\ref{eq:strange}, one should
quadratically add the following contributions: \\
-- the effect of the error of $\gep$ (Fig.~\ref{fig:proton}) \\
-- the effect of the error of $\gmp$ (Fig.~\ref{fig:proton}) \\
-- the effect of the error of $\gen$ (Fig.~\ref{fig:gen}) \\
-- the effect of the error of $\gmn$ (1.5\%) \\
-- the effect of the error of the  $e$--$p$ cross section (the denominator of
Eqs.~\ref{eq:born}-\ref{eq:strange}) (Fig.~\ref{fig:sigr}) \\
-- the uncertainty associated with neglected TPE corrections \\

Note that in evaluating the error due to $\gep$ and $\gmp$, the values of
the form factors are changed only in the numerator of Eq.~\ref{eq:born};
the value of $\sigr$ is left unchanged, as it's contribution to the uncertainty
is treated separately (Sec.~\ref{sec:sigr}). For the complete analysis of the
uncertainly of PV experiments, one must of course add the uncertainties
stemming from uncertainty in $\theta_W$ and $\gaz$ as well as uncertainty in
the scattering kinematics.

Finally, one obtains the term involving the strange form factors
by equating the term in Eq.~\ref{eq:strange} with ($\apv - \ans$).  Thus,
the uncertainty in $\sigr$ enters again when isolating the linear combination
$\ges + \eta \gms$.  Because $\apv \approx \ans$, the 1--2\% overall scale
uncertainty on the extracted value of $\ges + \eta \gms$ will always be very
small compared to the effect of the uncertainty of $\sigr$ on $\ans$, and so
again these uncertainties can be treated as uncorrelated without significant
effect on the final uncertainties.

\section{Conclusions}

We have evaluated the effect of TPE corrections on parity-violating elastic
electron--proton scattering using the TPE exchange calculations of
Refs.~\cite{blunden05a, borisyuk06}. The direct effect of TPE on the parity
violating asymmetry is small, $\ltorder$1\% for $Q < 1$~GeV/c. However,
the effect of TPE on the Rosenbluth extractions of $\gep$ and $\gmp$, which
are needed to extract the strangeness contribution from the asymmetry, can be
significant, and should be taken into account in the analysis of the
parity-violating measurements.  We have provided fits to the form factors and
their uncertainties, and provided a prescription to allows for an extraction
of the strangeness form factors without explicitly requiring a calculation of
the TPE exchange effects.  As we have shown, this prescription provides an
excellent approximation to the full procedure, based on tests performed using
the full TPE calculation.  This provides a common set of form factors and
uncertainties for the analysis of low-$Q$ parity violating measurements, as
well as a consistent application of TPE corrections on the extraction of the
strangeness form factors.  This approach has the advantage that it can be
applied without requiring an explicit calculation of the TPE amplitudes,
while  providing an approximation of the the TPE corrections to better than
0.2\%, with model dependence in the TPE correction that is consistent with the
assumed uncertainties due to RC for the measurements.

These TPE-corrected form factors are needed for determining the value and
uncertainty in $\apv$ for the case of no strange quark contributions.  They
are also the true, Born form factors that are related to the structure of
the nucleon, and which should be used in the analysis of other experiments.
However, the effect of TPE is different in different observables, and one
must consider if the Born form factors are the correct input in the case
being considered. For example, many analyses such as Rosenbluth separations
in quasielastic A(e,e'p) scattering~\cite{dutta03} or the extraction of $\gmn$
from polarization~\cite{anderson06} or ratio measurements~\cite{kubon02},
require knowledge of the $e$--$p$ cross section to extract information on other
quantities.  If one uses the TPE-corrected form factors, then one must include
TPE corrections in calculating the cross section.  In such cases, it is simpler
and less model-dependent to use the fits to the uncorrected cross sections.
Other cases, such as quasielastic neutrino scattering or the case of
parity-violating electron scattering considered here, will not have the same
TPE effects and need to be evaluated with care.

\begin{acknowledgments}

The authors thank P. Blunden and A. Kobushkin for providing calculations of
the two-photon amplitudes, and A. Afanasev, J. Jourdan, W. Melnitchouk, K.
Paschke, and D. Trautmann for useful discussions.  This work was supported by
the U. S. Department of Energy, Office of Nuclear Physics, under contract
W-31-109-ENG-38.

\end{acknowledgments}

\bibliography{hap}

\end{document}